\documentclass[12pt]{article}
\usepackage{graphics}
\usepackage{natbib}
\usepackage{url}
\usepackage{color}
\usepackage{epstopdf}
\usepackage{amsmath}
\usepackage{amssymb}
\usepackage{booktabs}
\usepackage{graphicx}
\usepackage{xcolor,colortbl}
\usepackage{lipsum}
\usepackage{setspace}
\usepackage{natbib}
\usepackage{multirow}
\usepackage{bm}
\usepackage{environ}
\usepackage{graphicx}
\usepackage{float}
\usepackage{textcomp}
\usepackage{comment}
\usepackage[pagewise]{lineno}

\renewcommand{\baselinestretch}{1.7}

\NewEnviron{NORMAL}{%
    \scalebox{2}{$\BODY$} 
} 

\NewEnviron{SMALL}{%
    \scalebox{0.5}{$\BODY$} 
} 

\begin{document}

\noindent{\Large {\bf Assessing the commonly used assumptions in estimating the principal causal effect in clinical trials}}

\vspace{0.1in}

\noindent
{\sl Yongming Qu$^{1\dag}$, Ilya Lipkovich$^1$, Stephen J. Ruberg$^2$}
\vspace*{0.1in}

{\footnotesize
\noindent
$^1$ Department of Data and Analytics, Eli Lilly and Company, Indianapolis, Indiana, 46285, USA\\
$^2$ Analytix Thinking, LCC, 11121 Bentgrass Court, Indianapolis, IN 46236, USA\\
$^\dag$Correspondce: Yongming Qu, Department of Data and Analytics, Eli Lilly and Company, Indianapolis, Indiana, 46285, USA. Email: qu\_yongming@lilly.com
}

\vspace{0.3in}
\newpage


\begin{abstract}
In clinical trials, it is often of interest to understand the principal causal effect (PCE), the average treatment effect for a principal stratum (a subset of patients defined by the potential outcomes of one or more post-baseline variables). Commonly used assumptions include monotonicity, principal ignorability, and cross-world assumptions of principal ignorability and  principal strata independence. In this article, we evaluate these assumptions through a 2$\times$2 cross-over study in which the potential outcomes under both treatments can be observed, provided there are no carry-over and study period effects. From this example, it seemed the monotonicity assumption and the within-treatment principal ignorability assumptions did not hold well. On the other hand, the assumptions of cross-world principal ignorability and cross-world principal stratum independence conditional on baseline covariates seemed reasonable. With the latter assumptions, we estimated the PCEs, defined by whether the blood glucose standard deviation increased in each treatment period, without relying on the cross-over feature, producing estimates close to the results when exploiting the cross-over feature. To the best of our knowledge, this article is the first attempt to evaluate the plausibility of commonly used assumptions for estimating PCEs using a cross-over trial. 
\end{abstract}

\newpage
\section{Introduction}
The concept of principal stratification was first introduced in the late 1990s and early 2000s, primarily in estimating the average treatment effect (ATE) for patients who can be compliant with their assigned treatment \citep{air1996identification,imbens1997bayesian,frangakis2002principal,hill2002differential}. 
A principal stratum is defined by the potential outcomes of one or more post-baseline variables, which generally cannot be observed for all patients in a parallel clinical trial, where each patient is randomized only to one of several candidate treatments. The post-baseline variable defining the principal stratum can be treatment compliance or adherence, occurrence of an adverse event, survivor's status, or biomarker response status. The estimation of the ATE for a principal stratum, also called principal causal effect (PCE), has attracted the interest of clinical trialists in recent years \citep{permutt2016taxonomy, permutt2018effects, qu2021implementation, bornkamp2021principal}. Most notably, the newly published \cite{international2020harmonised} listed principal stratification as one of the key strategies for defining estimands in clinical trials. 

Many approaches have been proposed for estimating PCEs, all requiring relatively strong assumptions, which are often untestable. One stream of research focuses on estimating PCEs via introducing some sensitivity parameters (not estimable from data alone) or estimating the bounds of PCEs \citep{mehrotra2006comparison, imai2008sharp,shepherd2011sensitivity,lou2019estimation}. The second line of research provides estimators based on the principal score, the probability of a subject belonging to the principal stratum as a function of baseline covariates \citep{hill2002differential, hayden2005estimator,jo2009use, ding2011identifiability,  feller2017principal,joffe2007defining, ding2017principal, small2017instrumental, bornkamp2020estimating, jiang2020multiply}. More recent research models the principal score indirectly via intermediate outcomes \citep{qu2020general}. The third line of research models principal strata effects within the mixture framework treating the observed data as mixtures of latent strata, typically employing Bayesian inference (\cite{imbens1997bayesian, magnusson2019bayesian}). Many approaches across all the lines rely on the {\em monotonicity} assumption to identify strata-specific effects, which are used in some of the research aforementioned and  others \citep{zhang2003estimation,jin2008principal,ding2011identifiability,chiba2011simple, vanderweele2011principal,lu2013rank,ding2017principal, magnusson2019bayesian,bornkamp2020estimating,jiang2020multiply}. 
Monotonicity basically assumes that a patient with a stratum-defining ``event" under one treatment would also have the same type of event under the alternative treatment.

The validity of the monotonicity and propensity score related assumptions are difficult to verify, especially in parallel clinical trials where each subject has the opportunity to receive only one treatment. There is little work in the literature to assess the validity of these assumptions in analyzing clinical trial data. To understand the plausibility of these assumptions, we looked into a 2$\times$2 cross-over clinical trial evaluating insulin treatments in patients with type-1 diabetes (expecting little carry-over and period effects) where each patient sequentially received both candidate treatments and therefore the principal strata could be naturally observed. We understand it is not possible to observe $\{A(0), Y(0)\}$ and $\{A(1), Y(1)\}$ simultaneously even in the cross-over study because in the cross-over study, the measurements $A$ and $Y$ under the two treatments are not observed at the exact same time and under the exact same condition; however, for a study with little carry-over and period effect, the observed outcomes for $A$ and $Y$ at each treatment period can be considered an approximation for $\{A(0),Y(0)\}$ and $\{A(1), A(1)\}$.  In this article, we share our findings, which we believe provide insight and directions for research into the estimation of PCEs. 

This article is organized as follows. In Section \ref{sec:methods} we review the commonly used assumptions for the estimation of the PCE. In Section \ref{sec:example} we will evaluate these assumptions through a 2$\times$2 cross-over study assuming no carry-over and period effect. Finally, Section \ref{sec:discussion} contains the summary and discussion. 

\section{Methods} \label{sec:methods}
We start by introducing the notation based on the counterfactual framework \citep{neyman1923, rubin1974}. Let $j=1,2,\ldots,n$ denote the index for $n$ subjects in a clinical trial (a finite population. For subject $j$, let $T_j$ denote the treatment indicator ($T_j=0$ for the control treatment and $T_j=1$ for the experimental treatment), $A_j$ be a post-baseline indicator variable to define the stratum of interest, $X_j$ be a vector of the baseline covariates, and  $Y_j$ denote the ultimate response variable. Additionally, we let $R_j$ denote the indicator of whether the ultimate outcome $Y_j$ is observed ($R_j=1$ for observed $Y_j$ and $R_j=0$ for missingness). To simplify the notation, we may drop the subscript $j$ if it does not cause confusion. Note, $A$ can be the treatment adherence indicator, an intercurrent events, or an indicator variable derived from one or more efficacy or safety variable(s).

The potential outcome under a treatment $t$ ($t=0,1$) is denoted by placing the treatment as an argument, ``$(t)$", following the variable name. For example, $Y_j(0)$ denotes the potential outcome for the response variable for subject $j$ under the control treatment. We consider principal strata $S_{kl}$ ($k=0,1,*$, $l=0,1,*$) defined by the potential outcomes $A(0)$ and $A(1)$:
 \begin{eqnarray}
 S_{kl} &=& \{j: A_j(0)=k, A_j(1) =l \}; \; k,l \in \{0,1\} \nonumber \\
 S_{k*} &=& \{j: A_j(0)=k \}, \label{eq:strata}\\
 S_{*l} &=& \{j: A_j(1) =l \}. \nonumber
 \end{eqnarray}
Note $S_{k*} = S_{k0} \cup S_{k1}$ and $S_{*l} = S_{0l} \cup S_{1l}$. 
Then, the PCE for the principal stratum $S_{kl}$ is defined as:
$$
\mu_{\mbox{\tiny PCE},kl} = |S_{kl}|^{-1} \sum_{j \in S_{kl}} E[Y_j(1) - Y_j(0) | S_{kl}],
$$
where $|S_{kl}|$ is the size of the set $S_{kl}$. For simplicity, we omit the summation sign and the subscript $j$ and write the ATE as:
\begin{equation} \label{eq:estimand_ps}
\mu_{\mbox{\tiny PCE},kl} = E[Y(1) - Y(0) | S_{kl}].
\end{equation}

Notable examples from causal literature and some recent contributions have been focusing on the PCEs for the principal strata of $S_{*1}$ and $S_{11}$, including:
\begin{enumerate}
    \item[$\bullet$] The complier average causal effect (CACE) for patients who can be compliant to a treatment \citep{air1996identification,imbens1997bayesian}.
    \item[$\bullet$] Survivor average causal effect (SACE) for patients who can be survivors under one or both treatments \citep{zhang2003estimation}.
    \item[$\bullet$] Causal effects on viral load for patients in vaccine trials who would be infected under one or both treatments \citep{gilbert2003sensitivity}.
    \item[$\bullet$] Adherer average causal effect (AdACE) for patients who can adhere to one or both treatments \citep{qu2020general}.
\end{enumerate}

The PCE in (\ref{eq:estimand_ps}) is not easy to estimate since the principal stratum is generally not observed for all patients. Additional assumptions are therefore required. A set of basic assumptions that are used in most causal inference frameworks is 
the {\em stable unit treatment value assumptions} (SUTVA) described in \cite{rubin1980randomization}:
\begin{flalign}
\quad\quad &\mbox{A1 } (\emph{SUTVA}): \; \left\{
\begin{array}{c} Y = Y(1)T+Y(0)(1-T) \\ 
A = A(1)T + A(0)(1-T). \end{array} \right. &&\nonumber
\end{flalign}
SUTVA implies the hypothetical outcome under a treatment is the same as the observed outcome under the same treatment, known as ``consistency'', implying ``no interference'' (treatment status of any patient does not affect the potential outcomes of other patients) and ``no-multiple-versions-of-treatment'' assumptions  \citep{vanderweele2013causal}.

Some literature discussed the use of the intermediate or ancillary outcomes in modeling $A$ or the potential outcome under the alternative hypothetical treatment \citep{qu2020general, louizos2017causal}; However, since the intermediate/ancillary potential outcomes under the alternative treatment are not observed, ultimately $A$ needs to be modeled through treatment and baseline covariates $X$. Therefore, we only consider modeling $A$ through $X$ and $T$ here.  
Three commonly used assumptions are:
\begin{flalign}
&\quad\quad \mbox{A2 (\emph{treatment ignorability})}: \; T \perp \{Y(0), Y(1), A(0), A(1) \} | X,  && \nonumber\\
&\quad\quad \mbox{A3 (\emph{principal ignorability})}: \; Y(t) \perp \{A(0), A(1)\} | X, \;\; \forall t=0,1, && \nonumber\\
&\quad\quad \mbox{A4 (\emph{explainable nonrandom principal strata})}: \; A(t) \perp \{Y(1-t), A(1-t)\} | X, \;\; \forall t=0,1. && \nonumber
\end{flalign}

Assumption A2 is generally called {\em treatment ignorability} assumption, Assumption A3 is the strong version of the \emph{principal ignorability} assumption, and Assumption A4 is also called the \emph{explainable nonrandom survival} assumption \citep{robins1998correction, hayden2005estimator}. Assumption A2 holds for all randomized clinical trials. Note Assumptions A3 and A4 have some overlap. 
To avoid overlapping assumptions, we define assumptions:
\begin{flalign}
&\quad\quad \mbox{A3$'$ (\emph{within-treatment principal ignorability})}: \; Y(t) \perp  A(t) | X_, \;\; \forall t=0,1,  && \nonumber\\
&\quad\quad \mbox{A3$''$ (\emph{cross-world principal ignorability})}: \; Y(t) \perp A(1-t)|X, \;\; \forall t=0,1, && \nonumber \\
&\quad\quad \mbox{A4$'$ (\emph{conditional cross-world principal stratum independence})}: \; A(0) \perp  A(1) | X. && \nonumber
\end{flalign}

We call A3$'$ the \emph{within-treatment principal ignorability} assumption, A3$''$ the \emph{cross-world principal ignorability} assumption, and A4$'$ the \emph{cross-world principal stratum conditional independence} assumption. Combined, Assumptions A3$''$ and A4$'$ are equivalent to Assumption A4 (due to the categorical nature of the variable $A$), but A3 is a sufficient but not necessary condition for A3$'$ and A3$''$. We also define the  (unconditional) \emph{cross-world principal stratum independence} assumption:
\begin{flalign}
&\quad\quad \mbox{A4$''$ (\emph{cross-world principal stratum independence})}: \; A(0) \perp  A(1). && \nonumber
\end{flalign}

With Assumptions A1, A2, and A3, a consistent estimator of the average potential outcome for $Y$ in the principal stratum strata $S_{k*}$ and $S_{*l}$ is provided in a few articles \citep{jo2009use, feller2017principal, bornkamp2020estimating}. \cite{hayden2005estimator} constructed estimators for principal stratum $S_{kl}$ ($k=0,1; l=0,1$) with the Assumptions A1, A2, and A4. Of note, in most references, the principal ignorability assumption only includes the independence of $A_j(t)$ and $Y_j(1-t)$ given $X_j$. While potential outcomes evaluated on the same patient for alternative treatments $t$ and $1-t$ are naturally correlated, it is not unreasonable to assume they are conditionally independent, given measured covariates.

Assumption A3$'$ is not relevant for the estimation of PCEs because the pair $\{Y(t), A(t)\}$ is either always observable (for the treatment actually assigned) or unobserved (for the treatment not assigned) in parallel studies, unless there are missing values (either due to true missingness or as a result of handling intercurrent events using a hypothetical strategy). In other words, there is no need to separate $\{Y(t), A(t)\}$; hence, whether $A(t)$ and $Y(t)$ are independent is not important. Sometimes additional assumptions are  required for estimating PCEs in the presence of missing values. For example, if $A$ is adherence status and non-adherent patients may discontinue from the study, such patients would have missing values for $Y$. Note, some non-adherent patients may remain in the study and have observed values for $Y$. How such values are used in analyses depends on the strategies used for handling intercurrent events in defining estimands as described in \cite{international2020harmonised}.
Also, sometimes $R=A$ or $R=1-A$ (e.g., $A$ is the indicator for survival), we should still use a different notation for the principal stratum variable and missingness indicator following the spirit of ICH E9 (R1). Detailed discussion and evaluation of the missingness assumptions are outside the scope of this article.

In addition to the assumptions for the (conditional) independence of $A$ and $Y$, one commonly used assumption is the {monotonicity} assumption:
\begin{flalign}
&\quad\quad \mbox{A5 (\emph{monotonicity})}: \; A(1) \ge A(0) \;  \text{ or } \; A(1) \le A(0). && \nonumber
\end{flalign}
The monotonicity assumption is commonly used with other assumptions to estimate PCEs or the upper or lower bounds of PCEs \citep{ zhang2003estimation,jin2008principal,magnusson2019bayesian,chiba2011simple, ding2011identifiability, vanderweele2011principal,  lu2013rank, bornkamp2020estimating}. One drawback for Assumption A5 is it ignores the stochastic nature of the random variables $A(t)$ and assumes a deterministic relationship between $A(1)$ and $A(0)$. 

For randomized parallel clinical trials, A2 can be guaranteed by randomization and A3$'$ may be evaluated using observed data, but A3$''$, A4$'$, A4$''$, and A5 cannot be evaluated. In Section \ref{sec:example}, we will evaluate these assumptions through a 2$\times$2 cross-over study, assuming no carry-over and study period effects. 

\section{Evaluating the commonly used assumptions using a real-life data example} \label{sec:example}
In this section, we will explore the plausibility of some assumptions listed in Section \ref{sec:methods} using  a 2$\times$2 cross-over study comparing two regimens of the basal insulin peglispro (BIL): a fixed-time  regimen given daily in the evening ($T=0$) versus a variable-time regimen given in the morning on Mondays, Wednesdays, and Fridays, and in the evening on Tuesdays, Thursdays, Saturdays, and Sundays ($T=1$) for patients with Type 1 diabetes mellitus. The study consists of a lead-in period of a daily injection of BIL at a fixed time each day for 12 weeks, then randomized to 2$\times$2 cross-over treatment periods with the above 2 regimens with 12 weeks for each period. Since insulins are given to control glucose without modifying the underlying disease, and their effect from the treatment in the first period can be washed out in a few weeks during the second period of treatment, there is no need for a separate washout period.
After the lead-in period, 182 patients were randomized to either the fixed/variable-time treatment sequence ($N$=92) or the variable/fixed treatment sequence ($N$=90), and 165 patients completed both treatment periods. The baseline characteristics and the main study results are published in \cite{garg2016similar}.

{At baseline and Week 12, patients were required to perform the fasting glucose measurement daily before the morning meal in at least 7 days prior to the visit. The BGSD was the standard deviation of the fasting glucose measurements in the 7 days prior to the visit. We consider $\Delta$BGSD, the change in BGSD from randomization to Week 12, as the response variable $Y$.}

We consider 2 variables to define principal strata:
\begin{enumerate}
    \item  { The indicator for whether $\Delta$BGSD is greater than 0, denoted by $A = I(\Delta\text{BGSD}>0)$, where $I$ is the indicator function.} The strata defined by this principal stratum variable may be used to answer questions like ``what is ATE if patients have an increase (or decrease) in BGSD under one or both treatments?". {The variable for this principal stratum is highly correlated with the outcome. }
    \item The occurrence of the treatment-emergent adverse events related to infections and infestations (TEAE-II) during the 12-week treatment period. {This variable is considered unrelated to the study medication and the treatment regimen because in other large studies there was no evidence suggesting BIL affect infection rates}.
\end{enumerate}

In this study, we expect the carry-over effect was small as the time-action of daily basal insulin is at most a few weeks and the period effect is minimal, as the type 1 diabetes disease progression is very slow. {Table \ref{tab:smy_BGSD} shows the mean and standard deviation for the baseline BGSD and $\Delta$BGSD at 12 weeks by treatment sequence and treatment regimen/period. For $\Delta$BGSD, the period and sequence effects were not statistically significant (p=0.09 for the period effect and 0.90 for the sequence effect) in a mixed linear model with the effects of sequence, treatment, period, and baseline BGSD as a covariate. Although this is not the definite evidence for no period and carry-over effects, the lack of statistical significance shows at least the period and carry-over effect were not strong.}
Therefore, tor the rest of this section, we assume there were no carry-over and period effects. We will discuss  these assumptions further in Section~\ref{sec:discussion}.  
\subsection{Evaluation of the monotonicity assumption} \label{sec:monotonicity}
In the first analysis, we evaluate the principal stratum variable defined by $\Delta$BGSD. For illustration purposes, we only include patients who had non-missing values for $\Delta$BGSD under both treatment regimens ($n=87$ for the fixed/variable-time sequence and $n=76$ for the variable/fixed-time sequence) so that we can understand the ``true" principal strata. Overall, the mean (standard deviation [SD]) baseline BGSD was 41.3 (22.4) mg/dL. The mean (SD) change in $\Delta$BGSD was -3.4 (25.4) mg/dL for the fixed-time regimen and -1.7 (24.8) mg/dL for the variable-time regimen.

Since both treatment regimens used the same medication except for the dosing time, the variable-time regimen theoretically should have larger $\Delta$BGSD than the fixed-time regimen. In this case, the (deterministic version of) the monotonicity assumption is: 
\[
A_j(1) \ge A_j(0) \quad \Leftrightarrow \quad \Pr(S_{10}) = 0.
\]

Since the injection device and the drug are exactly the same in both treatment regimens, we can assume the occurrence of TEAE-II is not related to the treatment regimen. Therefore, for the principal stratum variable of TEAE-II, the monotonicity assumption implies
\[
A(1) \ge A(0)   \mbox{ and } A(0) \ge A(1) \quad \Leftrightarrow \quad A(1)=A(0) \quad \Leftrightarrow \quad \Pr(S_{10}) = \Pr(S_{01}) = 0.
\]

Table \ref{tab:smy_ps} summarized the number (\%) of patients in  principal strata defined in Equation (\ref{eq:strata}) based on $\Delta$BGSD and TEAE-II. The shaded cells should have probability of 0 based on the monotonicity assumption described earlier in this section. However, for principal stratum variable defined by BGSD, the observed probability for $S_{01}$ was 19.0\%, which was much larger than 0. Similarly, for principal strata $S_{01}$ and $S_{10}$ defined by the adverse events of infections and infestations percentages, the observed probabilities were 16.4\% and 12.1\%, respectively, much larger than 0. In summary, with the assumptions of no carry-over and no period effects, it seemed the monotonicity assumption did not hold well for the two principal stratum variables.  

\subsection{Evaluating the principal ignorability assumption} \label{sec:ps_ignorability}
In this section, we evaluate assumptions A3$'$ and A3$''$ through estimating the mean response for $\Delta$BGSD for principal strata defined by the variable $A=I(\Delta\text{BGSD}>0)$. To assess A3$'$, we regress $Y(t)$ on $A(t)$ and $X$, with the adjustment for the potential (cross-over study) period effect. The inclusion of the period effect in the model is to evaluate if the assumption of no period effect is reasonable. If the assumption A3$'$ is valid, the coefficient for $A(t)$ should be close to 0. To assess A3$''$, we regress $Y(t)$ on $A(1-t)$ and $X$, with the adjustment for the period effect and assess the coefficient on $A(1-t)$. 

In all regressions, the period effect was not significant ($p > 0.2$) and the effect for the baseline covariate was significant ($p < 0.0001$). The estimated regression coefficients on $A(t)$ as well as the mean responses in strata $\{A(t)=0\}$ and $\{A(t)=1\}$ are provided in Table \ref{tab:reg_YAX}. The regression coefficients for $A(t)$ were highly significant when regressing $Y(t)$ on $A(t)$ and X. As a result, the estimate for $E\{Y(t)|\bar X, A(t)=0\}$ was very different from 
$E\{Y(t)|\bar X, A(t)=1\}$, where $\bar X$ is the mean for the baseline covariates. These results indicated Assumption A3$'$ was severely violated. This is not surprising because the stratum variable $A$ is defined by whether the change in BGSD from randomization is greater than zero, while $Y$ is the change in BGSD from randomization. On the other hand, the coefficient of $A(1-t)$ in the regression of $Y(t)$ on $A(1-t)$ and $X$ was not significant and the estimates for $E\{Y(t)|\bar X, A(1-t)=0\}$ and $E\{Y(t)|\bar X, A(1-t)=1\}$ were similar. This indicates that the cross-world Assumption A3$''$ is quite reasonable. 

{The above assessment for evaluating Assumptions A3$'$ and A3$''$ was based on linear models, which could have been misspecified. Nevertheless, if the conditional expectations between the 2 strata are very different, there is a strong evidence that the corresponding conditional independence assumption may be violated; however, on the other hand, when the conditional expectations were similar, it may not guarantee that the corresponding conditional independence assumption is satisficed. }


\subsection{Evaluating the cross-world principal stratum independence assumption} \label{sec:ps_cross_ind}

For most efficacy outcomes, the best baseline predictor of the post-baseline changes in outcome is the baseline score. Therefore, we model the principal score for the indicator of BGSD increase via baseline BGSD. Let $g(X_j,t)=E(A_j(t)|X_j)$ can be modeled through a logistic regression model with
\begin{equation}\label{eq:g}
g(X_j,t; \alpha_{0t}, \alpha_{1t}) = \frac{\exp(\alpha_{0t} + \alpha_{1t} X_j)}{1+\exp(\alpha_{0t} + \alpha_{1t} X_j)},
\end{equation}
where $X_j$ is the baseline BGSD. Assumption A4$'$ is equivalent to $$\Pr\{A_j(0)=k, A_j(1)=l|X_j\}=\Pr\{A_j(0)=k|X_j\}\Pr\{A_j(1)=l|X_j\}.$$
It implies
$$
\Pr\{A_j(0)=k, A_j(1)=l\} = E[\Pr\{A_j(0)=k|X_j\}\Pr\{A_j(1)=l|X_j\}].
$$
The left-hand quantity can be estimated by the mean of the observed values in $Y$ for observed principal strata when using the feature of the cross-over study (i.e. the availability of potential outcomes under both candidate treatments for the same patient) under the assumption of no carry-over and period effects. The right-hand quantity can be estimated as
$$
\hat E[\Pr\{A_j(0)=k|X_j\}\Pr\{A_j(1)=l|X_j\}] = n^{-1} \sum_{j=1}^n g(X_j,0,  \hat \alpha_{00}, \hat \alpha_{10}) g(X_j, 1,  \hat \alpha_{01}, \hat \alpha_{11}),
$$
where $\hat \alpha_{00}$ and $\hat \alpha_{10}$ are the parameter estimators from the logistic regression (\ref{eq:g}) only using data from the study periods with the control treatment, and $\hat \alpha_{00}$ and $\hat \alpha_{10}$ are the parameter estimators for the logistic regression (\ref{eq:g}) only using data from the study periods with the experimental treatment, without exploiting the crossing-over feature. 


Table \ref{tab:prob_stratum} shows the estimates of the probabilities of strata membership $S_{kl}$ assuming conditional independence A4$'$ without using the cross-over feature were very close to the observed proportions of patients in these strata when the cross-over feature is used. The standard errors in the parenthesis were estimated based on 5,000 bootstrap samples. On the other hand, the estimator  $\widehat{\Pr}\{A_j(0)=k\} \widehat{\Pr}\{A_j(1)=l\}$ assuming an unconditional cross-world independence A4$''$ provided results more different from the observed proportions compared to the estimates based on Assumption A4$'$, but did not reach significance ($p = 0.151$, calculated by the proportion of bootstrap samples have the same or more extreme differences between the two methods). These results suggest the conditional cross-world independence assumption A4$'$ is reasonable and the unconditional cross-world independence assumption may be a bit strong, but still reasonable for practical data analysis.

For the principal stratum variable of TEAE-II, there is no known baseline variable that can predict those adverse events. Therefore, we estimated the probability of patients in each stratum without including any baseline covariates (based on Assumption A4$''$). The estimates were significantly different from the observed proportions ($p = 0.038$).  This suggests that the unconditional principal stratum assumption A$4''$ is likely too strong. 


\subsection{Estimating the mean response for principal strata}
In this section, we estimated the mean $\Delta$BGSD for principal strata formed by $A = I(\Delta\mbox{BGSD}>0)$ without using the cross-over feature of the study. In Sections \ref{sec:monotonicity}, \ref{sec:ps_ignorability}, and \ref{sec:ps_cross_ind}, we showed the monotonicity assumption and within-treatment principal ignorability assumption were questionable, but the cross-world principal ignorability and the cross-world independence assumptions A3$''$ and A4$'$ seemed reasonable, conditional on baseline BGSD.  Therefore, we estimated the mean $\Delta$BGSD using the method provided by \cite{hayden2005estimator} based on Assumptions A3$''$ and A4$'$. The estimators for $\mu_{t,kl} := E\{Y(t)|S_{kl}\}$ are given by 
\begin{equation}\label{eq:mu_crover0}
\hat \mu_{0,kl} = \frac{\sum_{j=1}^n Y_j(0)A_j^k(0)\{1-A_j(0)\}^{1-k} \{g(X_j,1; \hat \alpha_{01}, \hat \alpha_{11})\}^l \{g(X_j,1; \hat \alpha_{01}, \hat \alpha_{11})\}^{1-l}}{\sum_{j=1}^n A_j^k(0)\{1-A_j(0)\}^{1-k} \{g(X_j,1; \hat \alpha_{01}, \hat \alpha_{11})\}^l \{g(X_j,1; \hat \alpha_{01}, \hat \alpha_{11})\}^{1-l} }
\end{equation}
and
\begin{equation}\label{eq:mu_crover1}
\hat \mu_{1,kl} = \frac{\sum_{j=1}^n Y_j(1)A_j^l(1)\{1-A_j(1)\}^{1-l} \{g(X_j,0; \hat \alpha_{00}, \hat \alpha_{10})\}^k \{g(X_j,0; \hat \alpha_{00}, \hat \alpha_{10})\}^{1-k}}{\sum_{j=1}^n A_j^l(1)\{1-A_j(1)\}^{1-l} \{g(X_j,0; \hat \alpha_{00}, \hat \alpha_{10})\}^k \{g(X_j,0; \hat \alpha_{00}, \hat \alpha_{10})\}^{1-k} }.
\end{equation}

We emphasize that equations (\ref{eq:mu_crover0}) and (\ref{eq:mu_crover1}) do not use the information on principal strata under alternative treatments observable only in a cross-over study (the cross-over feature of the trial). In other words, even though the principal score function is estimated using the same patients (hence not independent), we do not use the {\em paired} data $\{A_{j}(0), A_{j}(1)\}$. Instead, we use $A_{j}(0)$ and $A_{j}(1)$ in a {\em non-paired} fashion, similar to what we would have done in a parallel study. Although in this example both $Y_{j}(0)$ and $Y_{j}(1)$ were observed for the same subject $j$ (which obviously would not be the case in a parallel study), equations (\ref{eq:mu_crover0}) and (\ref{eq:mu_crover1}) do not require the $Y_{j}(0)$ and $Y_{j}(1)$ to be from the same subject as long as the corresponding principal scores can be estimated (which is true for parallel studies under Assumption A4).
Although we could easily derive the variance estimators for estimated principal strata effects in (\ref{eq:mu_crover0}) and (\ref{eq:mu_crover1}), we chose not to do so because the principal strata can be observed in a cross-over study and such a calculation has no practical use. For illustrative purposes, we calculate the 95\% confidence interval using the percentile bootstrap method. 

For this cross-over study, both $A_{j}(0)$ and ${A}_{j}(1)$ are observed, so a direct estimator of the mean response for each treatment arm in a stratum $S_{kl}$ $(k=0,1; l=0,1)$ can be obtained as a simple average of the outcomes for patients in that stratum:
\begin{equation}\label{eq:direct}
\widehat{\mu}_{t,kl}^{dir}= \frac{1}{|S_{kl}|} \sum_{\{j: j \in S_{kl}\}} Y_{j}(t),
\end{equation}
where $|S_{kl}|$ is the size of the set $S_{kl}$.

Table \ref{tab:ps} summarizes the mean $\Delta$BGSD for each treatment regimen and the treatment difference via the principal score and the direct estimation exploiting the cross-over study feature. The estimates (especially for the treatment difference) via the principal score were close to the direct estimates for all principal strata. This example suggests that the estimation of ATEs in principal strata via the principal score under the assumptions A3$''$ and A4$'$ is a viable approach in some situations. In addition, the results in Table \ref{tab:ps} shows the ATEs for different principal strata could be very different, indicating  the importance of considering the principal stratum. 

\section{Summary and Discussion} \label{sec:discussion}
In this article, we discussed the commonly used assumptions in estimating the PCEs and evaluated them through a 2$\times$2 cross-over design where the membership of patients in each principal stratum was naturally observed, assuming there were no carry-over and period effects, which are considered reasonable for this study. There was a 12-week lead-in period which eliminated a lot of noise, the time-action of daily basal insulin is at most a few weeks, and the type-1 diabetes disease progression is generally very slow. We considered a response variable of $\Delta$BGSD and 2 principal stratum variables: whether patients had $\Delta$BGSD $> 0$ and whether patients experienced TEAE-II in the analyses. 

The monotonicity assumption A5 has been widely used in the literature for estimating PCEs, but there has not been any evaluation of this assumption using real data. Analyses in this article based on data for several outcomes in a cross-over trial seemed not to support the monotonicity assumption. One may argue the observed outcomes at the end of each study period were not exactly the potential outcomes under treatments $t=0, 1$. While we agree on this argument, we have to acknowledge it is impossible to observe the perfect potential outcomes in any realistic setting. Heraclitus once said, ``no man ever steps in the same river twice, for it's not the same river and he's not the same man." A person stepping into the river twice between a short interval should serve a good approximation for ideal situation of the \emph{same} person steps in the \emph{same} river twice. Therefore, this 2$\times$2 cross-over study is the closest possible observable data to examine the assumption. The large proportions (greater than 10\%) for the cells that should have 0 probabilities based on the monotonicity assumption cannot be explained by the potential carry-over and period effects. In addition, by posing a deterministic condition on random variables $A(0)$ and $A(1)$, the monotonicity assumption seems to go against the statistical principles regarding randomness and variability. For example, a patient who discontinues their placebo treatment due to lack of efficacy may well have adhered to the experimental treatment for the duration of the study. Conversely, a patient who discontinues their experimental treatment due to an adverse event may well have adhered to the placebo treatment for the duration of the trial. With all said, we are not claiming the monotonicity is definitely invalid in this example or for any other cases. However, its nonstochastic nature, the lack of supportive empirical evidence in the literature, and insights from our analyses suggest the monotonicity is a rather strong assumption and one should take caution when using this assumption in estimating PCEs. 

We also found the within-treatment principal ignorability assumption A3$'$ to be overly optimistic. It seems more reasonable to assume the principal stratum depending on the observed postbaseline values \citep{qu2020general}. For example, if the principal stratum variable is the adherence status, patients experiencing serious AEs or lack of efficacy may be less likely to adhere to their assigned treatment. Note the assumption A3$'$ is analogous to the covariate-dependent missingness \citep{little1995modeling} in the literature on missing data. In clinical trials, missingness depending on the postbaseline observed values is generally considered a more reasonable assumption. In addition, A3$'$ is not an essential assumption for estimating PCEs. As explained in Section \ref{sec:methods}, assumptions on the mechanism of missingness (e.g., ignorable missingness), instead of the within-treatment principal ignorability, is needed to handle missing outcomes for the assigned treatment. 

Our analysis showed the cross-world principal ignorability assumption A3$''$ and the cross-world principal stratum independence assumption A4$'$ appeared reasonable when conditioned on baseline covariates. This finding is not surprising since the ``cross-worldness" will in general weaken the correlations between $A$ and $Y$. We estimated the PCE for stratum $S_{kl}$ under Assumptions A3$''$ and A4$'$ without using the cross-over feature. The results were very similar to the direct estimation of the corresponding PCE when exploiting the cross-over feature. This example suggests the estimator in \cite{hayden2005estimator} via principal scores with assumptions A3$''$ and A4$'$ may be reasonable to use in practice. 

{The PCE for adherers is an important and clinically meaningful estimand. However, we were not able to assess assumptions related to principal strata defined by adherence or compliance in this cross-over study. This is because we can only observe non-adherence under one treatment and the methods used in the assessment in this article require the observation of the potential outcomes under both treatments. Since patients may discontinue treatment and study in the first study period, we will never know the adherence status under the second study period under the alternative treatment. As there are not many cross-over studies in phase 2 and 3 development, we were only be able to assess the commonly used assumption for estimating PCEs in a single cross-over study. We welcome researchers to apply similar methods to other available cross-over studies.} 

In summary, in this article we used a 2$\times$2 cross study with expected weak or non-existent carry-over and period effects to gain insights into commonly used assumption in estimating PCEs. Our data analyses suggested the monotonicity and the within-treatment principal ignorablity assumptions were questionable, while the cross-world principal ignorability and cross-world principal stratum independence assumptions seemed reasonable. One limitation of this research is that we assumed no carry-over and study period effects. Nevertheless, we believe these findings will provide insights into the selection and development of statistical methods in estimating PCEs in the future.

\section*{Acknowledgements}
We would like to thank Yu Du and Dana Schamberger for their reviewing this manuscript and providing valuable comments. 




\bibliographystyle{apalike}
\bibliography{reference}

\renewcommand{\baselinestretch}{1.0}

\renewcommand{\baselinestretch}{1.0}
\begin{table}[htb]
\centering
\caption{The mean (standard deviation) of  BGSD at randomization and $\Delta$BGSD from baseline to 12 weeks.}
\label{tab:smy_BGSD}
\begin{tabular}{ccccc}
\hline\hline
Sequence     & Treatment (Period) && BGSD at Randomization & $\Delta$BGSD \\
     \cline{2-2} \cline{4-5}
\multirow{2}{*}{Fixed/Variable}
     &Fixed (Period 1) && \multirow{2}{*}{40.8 (23.7)} & -1.1 (24.0) \\
     &Variable (Period 2) &&   & -3.0 (25.9)\\
 \cline{2-5}
\multirow{2}{*}{Variable/Fixed}
     &Fixed (Period 2) && \multirow{2}{*}{41.9 (21.0)} & -5.9 (26.9) \\
     &Variable (Period 1) &&   & -0.1 (23.7)\\
\hline\hline
\end{tabular}
    {\begin{flushleft} Abbreviations: BGSD, the between-day fasting glucose variability measured by standard deviation; Fixed, fixed-time regimen; Variable, variable-time regimen. 
    \end{flushleft} }
\end{table}

\begin{table}[htb]
\centering
\caption{Number (\%) of patients by the principal stratum defined by $\Delta$BGSD}
\label{tab:smy_ps}
\begin{tabular}{ccccc}
\hline\hline
Principal Stratum Variable     && $A(1) = 0$ & $A(1) = 1$ & $A(1) \in \{0,1\}$ \\
     \cline{2-5}
\multirow{3}{*}{$\Delta$BGSD}
     &$A(0)=0$ & 52 (31.9) & 41 (25.2) & 93 (57.1)\\
     &$A(0)=1$ & \colorbox[gray]{0.8}{31 (19.0)} & 39 (23.9) & 70 (42.9)\\
     &$A(0) \in \{0,1\}$ & 83 (50.9) & 80 (49.1) & 163 (100)\\
 \cline{2-5}
\multirow{3}{*}{TEAE-II}
&$A(0)=0$ & 105 (63.6) & \colorbox[gray]{0.8}{20 (12.1)} & 125 (75.8)\\
&$A(0)=1$ & \colorbox[gray]{0.8}{27 (16.4)} & 13 (7.9) & 40 (24.2)\\
&$A(0) \in \{0,1\}$ & 132 (80.0) & 33 (20.0) & 165 (100)\\
\hline\hline
\end{tabular}
    {\begin{flushleft} Abbreviations: $\Delta$BGSD, the change in the between-day fasting glucose variability measured by standard deviation from randomization to Week 12; TEAE-II, treatment-emergent adverse events related to infections and infestations. 
    \end{flushleft} }
\end{table}

\begin{table}[htb]
\centering
\caption{The quantities estimated by regressing $Y$ on $A$ and $X$ with adjustment for the study period.}
\label{tab:reg_YAX}
\begin{tabular}{llcc}
\hline\hline
Regression & Quantity & Estimate (SE) & $p$-value \\
\cline{2-4}
\multirow{3}{*}{$Y(0)$ on $A(0)$ and $X$} & Coefficient for $A(0)$ & ~33.4 (2.6) & $<$.0001 \\
&$E\{Y(0)|\bar X_{\cdot}, A(0)=0\}$ & -17.7 (1.7) & --- \\
&$E\{Y(0)|\bar X_{\cdot}, A(0)=1\}$ & ~15.7 (1.9) & ---\\
\cline{2-4}
\multirow{3}{*}{$Y(1)$ on $A(1)$ and $X$} & Coefficient for $A(1)$ & ~31.5 (2.7) & $<$.0001 \\
&$E\{Y(1)|\bar X_{\cdot}, A(1)=0\}$ & -17.1 (1.8) &---\\
&$E\{Y(1)|\bar X_{\cdot}, A(1)=1\}$ & ~14.3 (1.9)& ---\\
\cline{2-4}
\multirow{3}{*}{$Y(0)$ on $A(1)$ and $X$} & Coefficient for $A(0)$ & ~1.3 (3.8) & 0.741 \\
&$E\{Y(0)|\bar X_{\cdot}, A(1)=0\}$ & -4.0 (2.6) & --- \\
&$E\{Y(0)|\bar X_{\cdot}, A(1)=1\}$ & -2.7 (2.6) & ---\\
\cline{2-4}
\multirow{3}{*}{$Y(1)$ on $A(0)$ and $X$} & Coefficient for $A(1)$ & ~3.6 (3.5) & 0.315 \\
&$E\{Y(1)|\bar X_{\cdot}, A(0)=0\}$ & -3.2 (2.3) & ---\\
&$E\{Y(1)|\bar X_{\cdot}, A(0)=1\}$ & ~0.3 (2.6) & ---\\
\hline\hline
\end{tabular}
\end{table}

\begin{table}[htb]
\centering
\caption{Estimated probability (standard error) for subjects belonging to principal strata $S_{kl}, (k,l\in\{0,1\}$) defined by the indicator of whether BGSD increased from baseline to 12 weeks}
\label{tab:prob_stratum}
\footnotesize
\begin{tabular}{cccccc}
\hline\hline
Stratum Variable & $\Pr(S_{kl})$ & $S_{00}$ & $S_{01}$ & $S_{10}$ & $S_{11}$ \\
\cline{2-6} 
&Observed & 0.319(0.037) & 0.252(0.034)& 0.190(0.031) & 0.239(0.033) \\
$\Delta$BGSD & Estimate assuming A4$'$ & 0.310(0.032) & 0.261(0.029) & 0.199(0.026) & 0.230(0.029) \\
& Estimate assuming A4$''$ & 0.291(0.032) & 0.280(0.027) & 0.219(0.024) & 0.211(0.027) \\
\cline{2-6}
\multirow{2}{*}{TEAE of II} &Observed & 0.636(0.037) & 0.121(0.025)& 0.164(0.029) & 0.079(0.021) \\
& Estimate assuming A4$''$ & 0.606(0.038) & 0.152(0.024) & 0.194(0.027) & 0.048(0.011) \\
\hline\hline
\end{tabular}
    {\begin{flushleft} Abbreviations: $\Delta$BGSD, the change in the between-day fasting glucose variability measured by standard deviation from randomization to Week 12; TEAE-II, treatment-emergent adverse events related to infections and infestations. 
    \end{flushleft} }
\end{table}

\begin{table}[htb]
\centering
\caption{Estimates for $\Delta$BGSD on principal strata defined by $A = I(\Delta\text{BGSD} > 0)$}
\label{tab:ps}
\begin{tabular}{ccccc}
\hline\hline
\multirow{2}{*}{Stratum} &  \multirow{2}{*}{Method} & \multicolumn{3}{c}{Mean (95\% confidence interval)}  \\
\cline{3-5}
 &  & FTR & VTR & FTR vs. VTR \\ \hline
\multirow{2}{*}{$S_{00}(n=52)$}  
         &DIRECT   & -22.2 (-27.4, -17.1)  & -20.3 (-25.4, -15.1)  &   -2.0 (-6.5,   2.6)\\
         &    PS   & -23.9 (-28.6, -19.4)  & -21.3 (-26.2, -16.8)  &   -2.7 (-7.0,   1.7)\\ \hline
\multirow{2}{*}{$S_{01}(n=41)$}
         &DIRECT   & -15.6 (-20.7, -10.5)  &  13.6 (8.9,  18.3)  &  -29.2 (-34.8, -23.7)\\
         &    PS   & -13.8 (-17.4, -10.7)  &  16.9 (13.3,  20.7)  &  -30.7 (-35.4, -26.2)\\ \hline
\multirow{2}{*}{$S_{10}(n=31)$}
         &DIRECT   &  15.4 (10.0,  20.8)  & -18.2 (-24.5, -12.0)  &   33.6 (25.6,  41.7)\\
         &    PS   &  18.0 (13.7,  22.8)  & -16.8 (-20.7, -13.5)  &   34.8 (29.4,  40.5)\\ \hline
\multirow{2}{*}{$S_{11}(n=39)$}                       
         &DIRECT   &  19.7 (13.4,  26.0)  &  20.1 (14.8,  25.5)  &   -0.42 (-8.4,   7.5)\\
         &    PS   &  17.6 (13.7,  22.0)  &  16.7 (13.3,  20.5)  &    0.94 (-4.3,   6.5)\\ 
\hline
\hline
\end{tabular}
    {\begin{flushleft} Abbreviations: $\Delta$BGSD, the change in the between-day fasting glucose variability measured by standard deviation from randomization to Week 12; FTR, fixed-time regimen; VTR, variable-time regimen; DIRECT, estimator directly based on the observed principal stratum using the cross-over feature; PS, estimator via the principal score.
    \end{flushleft} }
\end{table}
\label{lastpage}



\end{document}